# Social Cognitive Maps, Swarm Collective Perception and Distributed Search on Dynamic Landscapes

Vitorino Ramos[1], Carlos Fernandes[2,3], Agostinho C. Rosa[2,4]

[1] CVRM-IST, Technical Univ. of Lisbon (IST),
Av. Rovisco Pais, 1, 1049-001, Lisbon, PORTUGAL
`vitorino.ramos@alfa.ist.utl.pt`
[2] LaSEEB-ISR-IST, Technical Univ. of Lisbon (IST),
Av. Rovisco Pais, 1, TN 6.21, 1049-001, Lisbon, PORTUGAL
`{cfernandes,acrosa}@laseeb.org`
[3] EST-IPS, Setúbal Polytechnic Institute (IPS),
R. Vale de Chaves - Estefanilha, 2810, Setúbal, PORTUGAL
[4] Psycho Biological Dep., Univ. of São Paulo, São Paulo, SP, BRAZIL

**Abstract.** Swarm Intelligence (SI) is the property of a system whereby the collective behaviors of (unsophisticated) entities interacting locally with their environment cause coherent functional global patterns to emerge. SI provides a basis with which it is possible to explore collective (or distributed) problem solving without centralized control or the provision of a global model. To tackle the formation of a coherent social collective intelligence from individual behaviors, we discuss several concepts related to self-organization, stigmergy and social foraging in animals. Then, in a more abstract level we suggest and stress the role played not only by the environmental media as a driving force for societal learning, as well as by positive and negative feedbacks produced by the many interactions among agents. Finally, presenting a simple model based on the above features, we will address the collective adaptation of a social community to a cultural (environmental, contextual) or media informational dynamical landscape, represented here – for the purpose of different experiments – by several three-dimensional mathematical functions that suddenly change over time. Results indicate that the collective intelligence is able to cope and quickly adapt to unforeseen situations even when over the same cooperative foraging period, the community is requested to deal with two different and contradictory purposes.

## 1 Introduction

Swarm Intelligence (SI) is the property of a system whereby the collective behaviors of (unsophisticated) entities interacting locally with their environment cause coherent functional global patterns to emerge. SI provides a basis with which it is possible to explore collective (or distributed) problem solving without centralized control or the provision of a global model (Stan Franklin, *Coordination without Communication*, talk at Memphis Univ., USA, 1996). The well-know bio-inspired computational paradigms know as ACO (*Ant Colony Optimization* algorithm [8]) based on trail formation via pheromone deposition / evaporation, and PSO (*Particle Swarm Optimization* [24]) are just two among many successful examples. Yet, and in what specifically relates to the biomimicry of these and other computational models, much more can be of useful employ, namely the social foraging behavior theories of many species, which can provide us with consistent hints to algorithmic approaches for the construction of social cognitive maps, self-organization, coherent swarm perception and intelligent distributed search, with direct applications in a high variety of social sciences and engineering fields. In the present work, we will address the collective adaptation of a social community to a cultural (environmental, contextual) or informational dynamical landscape, represented here – for the purpose of different experiments – by several three-dimensional mathematical functions that change over time.

Flocks of migrating birds and schools of fish are familiar examples of spatial self-organized patterns formed by living organisms through social foraging. Such aggregation patterns are observed not only in colonies of organisms as simple as single-cell bacteria, as interesting as social insects like ants and termites as well as in colonies of multi-cellular vertebrates as complex as birds and fish but also in human societies [14]. Wasps, bees, ants and termites all make effective use of their environment and resources by displaying collective "swarm" intelligence. For example, termite colonies build nests with a complexity far beyond the comprehension of the individual termite, while ant colonies dynamically allocate labor to various vital tasks such as foraging or defense without any central decision-making ability [8,53]. Slime mould is another perfect example. These are very simple cellular organisms with limited motile and sensory capabilities, but in times of food shortage they aggregate to form a mobile slug capable of transporting the assembled individuals to a new feeding area. Should food shortage persist, they then form into a fruiting body that disperses their spores using the wind, thus ensuring the survival of the colony [30,44,53].

New research suggests that microbial life can be even richer: highly social, intricately networked, and teeming with interactions [47]. Bassler [3] and other researchers have determined that bacteria communicate using molecules comparable to pheromones. By tapping into this cell-to-cell network, microbes are able to collectively track changes in their environment, conspire with their own species, build mutually beneficial alliances with other types of bacteria, gain advantages over competitors, and communicate with their hosts - the sort of collective strategizing typically ascribed to bees, ants, and people, not to bacteria. Eshel Ben-Jacob [6] indicate that bacteria have developed intricate communication capabilities (e.g. quorum-sensing, chemotactic signalling and plasmid exchange) to cooperatively self-organize into highly structured colonies with elevated environmental adaptability, proposing that they maintain linguistic communication. Meaning-based communication permits colonial identity, intentional behavior (e.g. pheromone-based courtship for mating), purposeful alteration of colony structure (e.g. formation of fruiting bodies), decision-making (e.g. to sporulate) and the recognition and identification of other colonies – features we might begin to associate with a bacterial social intelligence. Such a social intelligence, should it exist, would require going beyond communication to encompass unknown additional intracellular processes to generate inheritable colonial memory and commonly shared genomic context. Moreover, Eshel [5,4] argues that colonies of bacteria are able to communicate and even alter their genetic makeup in response to environmental challenges, asserting that the lowly bacteria colony is capable of computing better than the best computers of our time, and attributes to them properties of creativity, intelligence, and even self-awareness.

These self-organizing distributed capabilities were also found in plants. Peak and co-workers [37,2] point out that plants may regulate their uptake and loss of gases by distributed computation – using information processing that involves communication between many interacting units (their *stomata*). As described by Ball [2], leaves have openings called stomata that open wide to let $CO_2$ in, but close up to prevent precious water vapour from escaping. Plants attempt to regulate their stomata to take in as much $CO_2$ as possible while losing the least amount of water. But they are limited in how well they can do this: leaves are often divided into patches where the stomata are either open or closed, which reduces the efficiency of $CO_2$ uptake. By studying the distributions of these patches of open and closed stomata in leaves of the cocklebur plant, Peak et al. [37] found specific patterns reminiscent of distributed computing. Patches of open or closed stomata sometimes move around a leaf at constant speed, for example. What's striking is that it is the same form of mechanism that is widely thought to regulate how ants forage. The signals that each ant sends out to other ants, by laying down chemical trails of pheromone, enable the ant community as a whole to find the most abundant food sources. Wilson [54] showed that ants emit specific pheromones and identified the chemicals, the glands that emitted them and even the fixed action responses to each of the various pheromones. He found that pheromones comprise a medium for communication among the ants, allowing fixed action collaboration, the result of which is a group behaviour that is adaptive where the individual's behaviours are not.

Some other authors also defend this self-organizing realm into brain function. As defended in [12], the self-organization of ants into a swarm and the self-organization of neurons into a brain-like structure are similar in many respects. As described earlier in here, swarms of social insects construct trails and networks of regular traffic via a process of pheromone laying and following. These patterns constitute what is known in brain science as a cognitive map. The main differences lies in the fact that insects write their spatial memories in the environment, while the mammalian cognitive map lies inside the brain, a detail that also constitutes an important advantage for artificial cognitive map formation in digital realms presented in past models [39]. As mentioned by the two authors, this analogy can be more than a poetic image, and can be further justified by a direct comparison with the neural processes associated with the construction of cognitive maps in the hippocampus. In [54], Wilson forecasted the eventual appearance of what he called "a stochastic theory of mass behavior" and asserted that "the reconstruction of mass behaviors from the behaviors of single colony members is the central problem of insect sociobiology". He forecasted that our understanding of individual insect behavior together with the sophistication with which we will able to analyze their collective interaction would advance to the point were we would one day posses a detailed, even quantitative, understanding of how individual "probability matrices" would lead to mass action at the level of the colony. As stated in [12], by replacing *colony members* with *neurons*, *mass behaviors* or *colony* by *brain behavior* and *insect sociobiology* with *brain science* the above paragraph could describe the paradigm shifts in the last twenty-five years of progress in the brain sciences [54] and Artificial Intelligence (AI) research. From traditional AI to cognitive sciences and connectionist models. In contrast with symbol processing and traditional AI approaches, in connectionist models representations of any phe-nomena are dynamically stored in the matrix of weights between pairs of processing elements and patterns to be retrieved are not explicitly stored anywhere. Retrieval is accomplished by activating a pattern of inputs, which is processed by the network of constituent elements, resulting in output of a recalled pattern. The answer is somewhere spread in the dynamics and interactions among several units of the system, not on the units itself, in conformity with Chris Langton's Artificial Life research area definition [26,40]. Either if these units are ants, bacteria, plants, fish, cells of an immune system or neurons. As recently stated by Chialvo [13,48], highly correlated brain dynamics produces synchronized states with no behavioral value, while weakly correlated dynamics prevents information flow. In [13], Chialvo discuss the idea put forward by Per Bak [1] that the working brain stays at an intermediate (critical) regime characterized by power-law correlations.

As early as 1979, Douglas Hofstadter [19,24] suggested that maybe the brain was like an ant colony. No single neuron in a brain contains knowledge in itself, it is only through their interaction that thinking can occur. As Hofstadter's Anteater explains:

*There is some degree of communication among the ants, just enough to keep them from wandering off completely at random. By this minimal communication they can remind each other that they are not alone but are cooperating with teammates. It takes a large number of ants, all reinforcing each other this way, to sustain any activity – such as trail building – for any length of time. Now my very hazy understanding of the operation of brain leads me to believe that something similar pertains to the firing of neurons...* (p. 316, [19]).

As acknowledged in [24], Hofstadter's view of ants is in line with the contemporary appreciation for the emergence of complex dynamical systems from the interactions of simple elements following simple rules. Like Hofstadter, Millonas [33,32,12] compares the communications network within a swarm of ants (pheromone-based and extremely dynamic) to the highly interconnected architecture of neurons in a brain. Both cases (*nodes* or *units* is in here an abstract term) can be described in terms of three characteristics: (1) their structure comprises a set of nodes and their interconnections, (2) the states of node variables change dynamically over time, and (3) there is learning – changes in the strengths of the connections among the nodes. An argument that goes in line with the famous Doyne Farmer's 1991 connectionism paper [16]. Millonas argues that the intelligence of an ant swarm arises during phase transitions – the same phase transitions that Langton described as defining the "edge of chaos" [27,26], or Wolfram's Class IV state [55], identifying a special dynamical regime – with open-ended novelty - squashed between chaotic regimes and the ones characterized by fixed points and cycle limits. Langton have suggested that the critical region between ordered and chaotic behaviour plays a central role in evolution, using a simple numerical parameter $\lambda$ for the measure of complexity in a CA (Cellular Automata). Systems that are in a state lying close to this edge of chaos (a critical value of $\lambda$) have the richest, most complex behaviour. This phase transition argument is also synchronized with the *Self-Organized Criticality* (SOC) theory of Per Bak [1], habitually recognized worldwide in his *sand pile* model, or in the frequency-size of earthquakes as well as city areas following a typical $1/f$ noise decay curve. As stated in [24], the movements of ants are essentially random as there is no systematic pheromone pattern; activity is a function of two parameters, which are the strength of the pheromone and the attractiveness of the pheromone to the ants. If the pheromone distribution is random or if the attraction of ants to the pheromone is weak, then no aggregation pattern will form. On the other hand, if a too-strong pheromone concentration is established, or if the attraction of ants to the pheromone is very intense, then a suboptimal pattern may emerge, as the ants crowd together in a sort of pointless conformity. At the edge, though, at the very edge of chaos were the parameters are tuned correctly, the ants will explore and follow the pheromone signals, and wander from the swarm, and come back to it, and eventually coalesce into a pattern that is, most of the time, the shortest, most efficient path from any given point to any other.

In these real and simulated examples of insect accomplishments, we see optimization of various types, whether clustering items [41,42] or finding the shortest path through a landscape, with certain interesting characteristics. None of these instances include global evaluation of the situation: an insect can only detect its immediate environment. In contrast, optimization traditionally requires some method for evaluating the fitness of a solution, which seems to require that candidate solutions be compared to some standard, which may be a desired goal state (e.g. classical Genetic Algorithms) or the fitness of other solutions (e.g. co-evolved Genetic Algorithms). The bottom-up methods of the insect societies however, permit no evaluation – no ant knows how well the swarm is doing. In general, the method of pheromone indirect communication means that a more successful path will be somewhat more attractive, with an autocatalytic accumulation of pheromone resulting in the population's convergence on the most-fit behaviour – all done at the local level [24].

If an ant colony on his cyclic way from the nest to a food source (and back again), has only two possible branches around an obstacle, one bigger and the other smaller (*the bridge experiment* [7,52]), pheromone will accumulate – as times passes – on the shorter path, simple because any ant that sets out on that path will return sooner, passing the same points more frequently, and via that way, reinforcing the signal of that precise branch. Even if as we know, the pheromone evaporation rate is the same in both branches, the longer branch will faster vanish his pheromone, since there is not enough critical mass of individuals to keep it. On the other hand – in what appears to be a vastly pedagogic trick of Mother Nature - evaporation plays a critical role on the society. Without it, the final global decision or the phase transition will never happen. Moreover, without it, the whole colony can never adapt if the environment suddenly changes (e.g., the appearance of a third even shorter branch). While pheromone reinforcement plays a role as system's memory, evaporation allows the system to adapt and dynamically decide, without any type of centralized or hierarchical control.

Similar social foraging capabilities have also been a source of inspiration to some authors in the area of distributed Optimization and Control. Based on the biology and physics underlying the foraging behaviour of *E. coli* bacteria, Passino and Liu [36, 28,29] exploit a variety of bacterial swarming and social foraging behaviours, discussing how the control system on the *E. coli* dictates how foraging should proceed. From here, an algorithmic model is presented that emulates the distributed optimization process of the entire colony, with applications to a simple multiple-extremum function minimization problem, as well as – in a brief final discussion – to develop adaptive controllers and cooperative control strategies for autonomous vehicles, a study later extended to noisy environments [29]. Other authors used similar bacterial colony approaches for the optimization of Mobile Robot path planning [46]. At a more theoretical level, Panait and Luke [35] presented recently a pheromone-based model for collaborative foraging, and in a very interesting previous work even studied how social foraging behaviours could computationally evolve [34].

Finally, other related paradigms and biology-inspired computing models [11] involve the research and developing of Artificial Immune Systems (AIS) [15,10]. The biological immune system is a highly parallel and distributed adaptive system. It uses learning, memory, and associative retrieval to solve recognition and classification tasks. In particular, it learns to recognize relevant patterns, remember patterns that have been seen previously, and use combinations to construct pattern detectors efficiently. These remarkable information-processing abilities of the immune system provide important aspects in the field of computation, as in the specific case of agent-based AIS systems provided by Grilo, Caetano and Rosa [18].

All these above mentioned aspects show how vital can be the study of social foraging for the development of new distributed search algorithms, and the construction of social cognitive maps, with interesting properties in collective memory, collective decision-making and swarm pattern detection. In the present work, we consider how mechanisms inspired by social insects, in particular chemical signaling, may be used to control large homogeneous populations of simple ant-like agents, so that the collective result of their individual behaviors is the achievement of a particular global goal. Specifically, the agents are required to move from a randomly distributed initial spatial configuration to a tightly constrained configuration, as a 3D landscape peak. In order to show that the system is highly robust as well extremely adaptive, we study the colony behavior when one landscape suddenly changes into another. To achieve it, we present a simple extended computer model based on pheromone distribution, which represents the memory of the recent history of the swarm, and in a sense it contains information which the individual ants are unable to hold or transmit. There is no direct communication between the organisms but a type of indirect communication through the *pheromonal* field. In this case, the environment plays a critical role on contextual learning among the collective. On the other hand, these artificial ants sense and act locally, so that control in the population as a whole is fully decentralized. Since, the main aspects relate to self-organization and stigmergy, we first introduce them in some detail as well as some universal guidelines that we consider relevant to the development of highly distributed swarm systems of any kind.

## 2 Self-Organization and Stigmergy

Many structures built by social insects are the outcome of a process of self-organization, in which the repeated actions of the insects in the colony interact over time with the changing physical environment to produce a characteristic end state [20]. A major mediating factor is stigmergy [52], the elicitation of specific environment-changing behaviors by the sensory effects of local environment changes produced by previous and past behavior of the whole community. Stigmergy is a class of mechanisms that mediate animal-animal interactions through artifacts or via indirect communication, providing a kind of environmental synergy, information gathered from work in progress, a distributed incremental learning and memory among the society. In fact, the work surface is not only where the constituent units meet each other and interact, as it is precisely where a dynamical cognitive map could be formed, allowing for the embodiment of adaptive memory, cooperative learning and perception [39]. Constituent units not only learn from the environment as they can change it over time. Its introduction in 1959 by Pierre-Paul Grassé[1] made it possible to explain what had been until then considered paradoxical observations: In an insect society individuals work as if they were alone while their collective activities appear to be coordinated. The stimulation of the workers by the very performances they have achieved is a significant one inducing accurate and adaptable response. The phrasing of his introduction of the term is worth noting (translated to English in [20]):

*The coordination of tasks and the regulation of constructions do not depend directly on the workers, but on the constructions themselves.* **The worker does not direct his work, but is guided by it**. *It is to this special form of stimulation that we give the name Stigmergy (**stigma** - wound from a pointed object, and **ergon** - work, product of labor = stimulating product of labor).*

## 2.1 The role of the Environment

These self-organizing complex systems typically are comprised of a large number of frequently similar components or events, occurring at a common environment. Trough their process, a behavioral or physical pattern at the global-level of complexity, emerges solely from numerous interactions. As provided by Bonabeau and colleagues [7,9], self-organization (hereafter SO) refers to a set of dynamical mechanisms whereby structures appear at the global level of a system from interactions among its lower-level components. Moreover, the rules specifying the interactions among the system's constituent units are executed on the basis of purely local information, without reference to the global pattern, which is an emergent property of the system rather than a property imposed upon the system by an external ordering influence. On the other hand, because the global (collective) properties of the system often defy intuitive understanding of their origins, those properties may seem to appear mysteriously. There is nothing mystical or unscientific about their emergence, however [9]. Stigmergy is in fact, a well know example of self-organization in biological systems, and can provide not only vital clues in order to understand how the components can interact to produce a complex pattern, as can pinpoint simple biological non-linear rules to achieve improved artificial intelligent systems, as those found within Swarm Intelligence [42,8]. A paradigmatic case is provided by the emergence of self-organization in social insects, by way of indirect interactions. An example could be provided by two individuals, who interact indirectly when one of them modifies the environment and the other responds to the new environment at a later time. Grassé showed that the coordination and regulation of building activities do not depend on the workers themselves but are mainly achieved by the nest structure: a stimulating configuration triggers the response of a termite worker, transforming the configuration into another configuration that may trigger in turn another (possibly different) action performed by the same termite or any other worker in the colony (*qualitative stigmergy*). Another well-known illustration of how stigmergy and self-organization can be combined into more subtle adaptive behaviors is recruitment in social insects. Self-organized trail laying via pheromone by individual ants (*quantitative stigmergy*) is a way of modifying the environment to communicate with nest mates that follow such trails [12,39]. It appears that task performance by some workers decreases the need for more task performance: for instance, nest cleaning by some workers reduces the need for nest cleaning [8]. Therefore, nest mates communicate to other nest mates by modifying the environment (cleaning the nest), and nest mates respond to the modified environment (by not engaging in nest cleaning); that is, stigmergy.

---

[1] Grassé, P.P.: La reconstruction du nid et les coordinations inter-individuelles chez *Bellicositermes natalensis* et *Cubitermes sp*. La théorie de la stigmergie : Essai d'interpretation des termites constructeurs. *Insect Sociaux* (1959), 6, 41-83.

| *A*    | *t*=1   | *t*=10   | *t*=30   | *t*=40   | *t*=50   | *t*=80   |
| *t*=100 | *t*=120 | *t*=150  | *t*=200  | *t*=500  | *t*=1000 | *t*=1500 |
| *t*=2000 | *t*=2500 | *t*=3000 | *t*=3500 | *t*=4000 | *t*=4500 | *t*=5000 |
| *t*=5500 | *t*=6000 | *B*      |          |          |          |          |

Fig.1 – One swarm (3000 ants) is thrown to explore *Kafka* digital image (**A**) for 6000 iterations. At *t*=100, the *Kafka* image *habitat* is replaced by *Red Ant* image (**B**). Evolutions of swarm cognitive maps (pheromonal fields) are shown for several iterations. Darker pixels correspond to higher concentrations of pheromone. Results from [39,40].

Division of labor is another paradigmatic phenomenon of stigmergy. Simultaneous task performance (parallelism) by specialized workers is believed to be more efficient than sequential task performance by unspecialized workers [21]. Parallelism avoids task switching, which costs energy and time. A key feature of division of labor is its plasticity, being rarely rigid [43]. The ratios of workers performing the different tasks that maintain the colony's viability and reproductive success can vary in response to internal perturbations or external challenges. In other words, local perception and action taken by each agent can evolve adaptive and flexible problem-solving mechanisms, or emerge communication among many parts, at the colony level.

Another crucial example is how ants form piles of items such as dead bodies (corpses), larvae, or grains of sand. There again, stigmergy is at work: ants deposit items at initially random locations. When other ants perceive deposited items, they are stimulated to deposit items next to them, being this type of cemetery clustering organization and brood sorting a type of self-organization and adaptive behavior. Models of nest building in wasps [50,51] for instance, were already described, in which wasp-like agents are stimulated to deposit bricks when they encounter specific configurations of bricks: depositing a brick modifies the environment and hence the stimulatory field of other agents. These asynchronous automata (designed by an ensemble of algorithms) move in a 3D discrete space and behave locally in space and time on a pure stimulus-response basis. There are other types of examples (e.g. prey collectively transport), yet stigmergy is also present: ants change the perceived environment of other ants (the colony cognitive map, according to [12,32,33]), and in every example, the environment serves as medium of communication.

## 2.2 The role of Positive and Negative Feedbacks

Being these systems comprised of a large number of frequently similar components or events, the principal challenge is to understand how the components interact to produce a complex pattern [9]. Understanding the nature and properties of Self-Organization (SO), and how it works, is not only crucial as can be of great insight. A suitable approach is to first understand the two basic modes of interaction among the components of SO systems: positive and negative feedback. Bonabeau et al [7,9] identified four basic ingredients (negative and positive feedbacks, the amplification of fluctuations, and the presence of multiple interactions) and three characteristic signatures (the creation of spatiotemporal structures in an initially homogeneous medium, the possible attainability of different stable states, i.e. multi-stability, and the existence of parametrically determined bifurcations). Let's now described the two most important in some detail:

- **Negative feedback**, $f$: a mechanism familiar to biologists, commonly used to stabilize physiological processes (homeostasis) and avoid undesirable fluctuations. A well-known example is the regulation of blood sugar levels, a process that proceeds smoothly in most people but functions abnormally in diabetics. Blood sugar levels are regulated by a negative feedback mechanism involving the release of insulin. An increase in blood glucose following ingestion of sugary meal quickly triggers the release of insulin from the pancreas, resulting in a number of physiological effects, including the counteraction of the increase in blood-sugar level. In this case, we see $f$ acting to maintain the status quo by damping large fluctuations in blood glucose level. A similar example involves the homeostatic regulation of body temperature. If the organism experiences a thermal stress, then a discrepancy between the organism's actual body temperature (monitored by sensitive receptors in the hypothalamic area of the brain) and his internal thermal set point, triggers various behavioral and physiological responses. In both instances, the individual acquires and processes information that elicits a negative feedback response, that is, a small perturbation applied to the system triggers an opposing response that counteracts the perturbation. This opposing response is normally set in action *globally* on the entire system. In the first case an *increase* in blood glucose triggers a compensating response leading to a *de-*

*crease* in blood glucose. In the second case, a *decrease* in body temperature results in responses that *increase* body temperature. In other cases, negative feedback plays a continuously role, without the necessity of having *increase* or *decrease* signals in order to trigger it. A well-known example is provided by pheromone evaporation on the self-organization of trails in some social insects, which acts globally and continuously over time. Without evaporation the colony use of pheromone is useless, since the whole system keeps reinforcing the same path (solution) over and over again, being extremely fragile – and not robust - to new situations (e.g. landscape sudden change). Other similar example is provided by people walking in footpaths, where physical or weather conditions, such as erosion, play the role of negative feedback. The overall final result however, is a dynamical robust memory [39,40] (a kind of *collective conscience*) shared by a community seeking to find the "right", appropriate or aesthetical path, for instance, between two villages through a dense forest. What strikes from this example, is that: 1) the community arrive at a common path, without direct communication, 2) the signal reinforcement (see positive feedback) is made discontinuously in time and, 3) at every new surprise (a landslide over parts of the trail, for instance), rapidly the distributed whole of people locally finds a new solution to re-connect the entire path again and again. As a mnemonically brain activation, a sort of environmental synergy. The entire social cognitive map as the sum of these distributed cues, is thus changing over time, and intelligently adapting, or self-organizing to new unforeseen environmental situations without a global master plan. As if the environment is cooperating directly with the whole un-corporated mental map, to form an entirely new, alive and embodied phenomenon. Without landslides or obstacles like falling trees and people "getting lost" finding new solutions, the path will never re-organize to explore new geographical areas, and without having people locally to reinforce cues, the path in the first place will never emerge. For better or for worse, without negative feedback the systems stables or freezes itself on a particular configuration, an entire trail that never changes, keeps reinforcing itself and will never evolve over time. Simultaneously and equally important, the SO system needs components (people, in this case) that reinforce previous solutions (*exploitation* of the search space) and components that more or less randomly, having or not a purpose to it, subject themselves to areas "off the beaten track" (*exploration*), ideal to attract - with their first and still fragile cues - the entirely community into a new global solution in case of a sudden change or crisis (landslides). Often, components of the SO system can do both, based on their non-linear response thresholds to environmental stimulus [41], as in the model presented later.

- **Positive feedback**, $f^+$: in contrast to negative feedback, positive feedback generally promotes changes in the system (the majority of SO systems use them). The explosive growth of the human population provides a familiar example of the effect of positive feedback. The snowballing autocatalytic effect of $f^+$ takes an initial change in a system (due to amplification of fluctuations; a minimal and natural local cluster of objects could be a starting point) and reinforces that change in the *same* direction as the initial deviation. Self-enhancement, amplification, facilitation, and autocatalysis are all terms used to describe positive feedback [9]. Another example could be provided by the clustering or aggregation of individuals. Many birds, such as seagulls nest in large colonies. Group nesting evidently provides individuals with certain benefits, such as better detection of predators or greater ease in finding food. The mechanism in this case is imitation[2]: birds preparing to nest are attracted to sites where other birds are already nesting, while the behavioral rule could be synthesized as "I nest close where you nest". The key point is that aggregation of nesting birds at a particular site is *not* purely a consequence of each bird being attracted to the site per se. Rather, the aggregation evidently arises primarily because each bird is attracted to *others* (check for further references on [7,9]). On social insect societies, $f^+$ could be illustrated by the pheromone reinforcement on trails, allowing the entire colony to exploit some past and present solutions. Generally, as in the above cases, positive feedback is imposed implicitly on the system and *locally* by each one of the constituent units. Fireflies flashing in synchrony [49] follow the rule, "I signal when you signal", fish traveling in schools abide by the rule, "I go where you go", and so forth. In humans, the "infectious" quality of a yawn of laughter is a familiar example of positive feedback of the form, "I do what you do". Seeing a person yawning[3], or even just thinking of yawning, can trigger a yawn [9]. There is however one associated risk, generally if $f^+$ acts alone without the presence of negative feedbacks, which *per si* can play a critical role keeping under control this snowballing effect, providing inhibition to offset the amplification and helping to shape it into a particular pattern. Indeed, the amplifying nature of $f^+$ means that it has the potential to produce destructive explosions or implosions in any process where it plays a role. Thus the behavioral rule may be more complicated than initially suggested, possessing both an autocatalytic as well as an antagonistic aspect. In the case of fish [9], the minimal behavioral rule could be "I nest where others nest, *unless the area is overcrowded*". In this case both the positive and negative feedback may be coded into the behavioral rules of the fish. Finally, in other cases one finds that the inhibition arises automatically, often simply from physical constraints.

Since in SO systems their organization arises entirely from multiple interactions, it is of critical importance to question how organisms acquire and act upon information [9]. Basically through two forms: *a*) information gathered from one's neighbors, and *b*) information gathered from work in progress, that is, stigmergy. In the case of animal groups, these internal interactions typically involve information transfers between individuals. Biologists have recently recognized that information can flow within groups via two distinct pathways – signals and cues. Signals are stimuli shaped by natural selection specifically to convey information, whereas cues are stimuli that convey information only incidentally [9]. The distinction between signals and cues is illustrated by the difference ant and deer trails. The chemical trail deposited by ants as they return from a desirable food source is a signal. Over evolutionary time such trails have been molded by natural selection for the purpose of sharing with nestmates information about the location of rich food sources. In contrast, the rutted trails made by deer walking through the woods is a cue, not shaped by natural selection for communication among deer but are a simple by-product of animals walking along the same path. SO systems are based on both, but whereas signals tends to be conspicuous, since natural selection has shaped signals to be strong and effective displays, information transfer via cues is often more subtle and based on incidental stimuli in an organism's social environment [45].

---

[2] See also on this subject the seminal sociological work of *Gabriel Tarde*; Tarde, G., *Les Lois de l'Imitation*, Eds. du Seuil (2001), 1st Edition – Eds. Alcan, Paris, 1890.

[3] Similarly, *Milgram* et al (Milgram, Bickerman and Berkowitz, "Note on the Drawing Power of Crowds of Different Size", *Journal of Personality and Social Psychology*, 13, 1969) found that if one person stood in a Manhattan street gazing at a sixth floor window, 20% of pedestrians looked up; if five people stood gazing, then 80% of people looked up.

## 3  A Swarm Cognitive Map Model for Dynamic Landscapes

As we shall see, the distribution of the pheromone represents the memory of the recent history of the swarm, and in a sense it contains information which the individual ants are unable to hold or transmit. There is no direct communication between the organisms but a type of indirect communication through the *pheromonal* field. In fact, ants are not allowed to have any memory and the individual's spatial knowledge is restricted to local information about the whole colony pheromone density. In order to design this behaviour, one simple model was adopted [12], and extended due to specific constraints of the present proposal, in order to deal with 3D dynamic landscapes. As described by *Chialvo* and *Millonas* in [12], the state of an individual ant can be expressed by its position $r$, and orientation $\theta$. Since the response at a given time is assumed to be independent of the previous history of the individual, it is sufficient to specify a transition probability from one place and orientation $(r,\theta)$ to the next $(r^*,\theta^*)$ an instant later. In previous works by Millonas [33,32], transition rules were derived and generalized from noisy response functions, which in turn were found to reproduce a number of experimental results with real ants. The response function can effectively be translated into a two-parameter transition rule between the cells by use of a pheromone weighting function (Eq.1):

$$W(\sigma) = \left(1 + \frac{\sigma}{1+\gamma\sigma}\right)^{\beta} \qquad (1)$$

This equation measures the relative probabilities of moving to a cite $r$ (in our context, to a cell) with pheromone density $\sigma(r)$. The parameter $\beta$ is associated with the osmotropotaxic sensitivity, recognised by Wilson [54] as one of two fundamental different types of ant's sense-data processing. *Osmotropotaxis*, is related to a kind of instantaneous pheromonal gradient following, while the other, *klinotaxis*, to a sequential method (though only the former will be considered in the present work as in [12]). Also it can be seen as a physiological inverse-noise parameter or gain. In practical terms, this parameter controls the degree of randomness with which each ant follows the gradient of pheromone. On the other hand, $1/\gamma$ is the sensory capacity, which describes the fact that each ant's ability to sense pheromone decreases somewhat at high concentrations.

| F0a - 3D view | F0a - 2D view | F0b - 3D view | F0b - 2D view |
| F1 - 3D view | F1 - 2D view | F2 - 3D view | F2 - 2D view |
| F3 - 3D view | F3 - 2D view | F4 - 3D view | F4 - 2D view |
| F6 - 2D view | F6 - 2D view | | |

Fig.2 – Three-dimensional views (3D) and respective landscapes views (2D) of several test functions used in our analysis [38]. White pixels correspond to high peaks, while darker ones represent deep valleys (*F0-F4*) or holes (*F6*). Check table 1 in section 4.

| $t=0$ | $t=0$ | $t=500$ | $t=500$ |
| $t=50$ | $t=50$ | $t=1000$ | $t=1000$ |
| $t=100$ | $t=100$ | | |

Fig.3 – **maxF0a**. Pheromone distribution (Social Cognitive Maps) for $t=0$, 50, 100, 500 and 1000 time steps, of 3000 ants exploring function *F0a* on a 100 x 100 toroidal grid (1$^{st}$ and 3$^{rd}$ column: darker pixels correspond to higher concentrations). Columns 2 and 4 correspond to the geographical place where agents are situated (each black pixel is an ant). At $t=100$, the highest peak is already surrounded by agents while convergence proceeds. Processing time equals to 54 *s*.

$$P_{ik} = \frac{W(\sigma_i)w(\Delta_i)}{\sum_{j/k} W(\sigma_j)w(\Delta_j)} \quad (2)$$

In addition to the former equation, there is a weighting factor $w(\Delta\theta)$, where $\Delta\theta$ is the change in direction at each time step, i.e. measures the magnitude of the difference in orientation. As an additional condition, each individual leaves a constant amount $\eta$ of pheromone at the cell in which it is located at every time step *t*. This pheromone decays at each time step at a rate *k*. Then, the normalised transition probabilities on the lattice to go from cell *k* to cell *i* are given by $P_{ik}$ (Eq. 2, [12]), where the notation *j/k* indicates the sum over all the surrounding cells *j* which are in the local neighbourhood of *k*. $\Delta_i$ measures the magnitude of the difference in orientation for the previous direction at time *t*-1. That is, since we use a neighbourhood composed of the cell and its eight neighbours, $\Delta_i$ can take the discrete values 0 through 4, and it is sufficient to assign a value $w_i$ for each of these changes of direction. Chialvo et al. used the weights of $w_0 =1$ (same direction), $w_1 =1/2$, $w_2 =1/4$, $w_3 =1/12$ and $w_4 =1/20$ (U-turn). In addition, coherent results were found for $\eta=0.07$ (pheromone deposition rate), $k=0.015$ (pheromone evaporation rate), $\beta=3.5$ (osmotropotaxic sensitivity) and $\gamma =0.2$ (inverse of sensory capacity), where the emergence of well defined networks of trails were possible. Except when indicated, these values will remain in the following framework. As an additional condition, each individual leaves a constant amount $\eta$ of pheromone at the cell in which it is located at every time step t. Simultaneously, the pheromone evaporates at rate $k$, i.e., the pheromonal field will contain information about past movements of the organisms, but not arbitrarily in the past, since the field *forgets* its distant history due to evaporation in a time $\tau \cong 1/k$. As in past works, toroidal boundary conditions are imposed on the lattice to remove, as far as possible any boundary effects (e.g. one ant going out of the grid at the south-west corner, will probably come in at the north-east corner).

In order to achieve emergent and *autocatalytic* mass behaviours around specific locations (e.g., peaks or valleys) on the *habitat*, instead of a constant pheromone deposition rate $\eta$ used in [12], a term not constant is included. This upgrade can significantly change the expected ant colony cognitive map (pheromonal field). The strategy follows an idea implemented earlier by Ramos [39] (fig. 1), while extending the Chialvo model into digital image habitats, aiming to achieve a collective perception of those images by the end product of swarm interactions. The main differences to the Chialvo work, is that ants now move on a 3D discrete grid, representing the functions which we aim to study (fig. 2) instead of a 2D *habitat*, and the pheromone update takes in account not only the local pheromone distribution as well as some characteristics of the cells around one ant. In here, this additional term should naturally be related with specific characteristics of cells around one ant, like their altitude ($z$ value or function value at coordinates $x,y$), due to our present aim. So, our pheromone deposition rate $T$, for a specific ant, at one specific cell $i$ (at time $t$), should change to a dynamic value ($p$ is a constant = 1.93) expressed by equation 3. In this equation, $\Delta_{max} = | z_{max} - z_{min} |$, being $z_{max}$ the maximum altitude found by the colony so far on the function *habitat*, and $z_{min}$ the lowest altitude. The other term $\Delta[i]$ is equivalent to (if our aim is to minimize any given landscape): $\Delta[i] = | z_i - z_{max} |$, being $z_i$ the current altitude of one ant at cell $i$. If on the contrary, our aim is to maximize any given landscape, then we should instead use $\Delta[i] = | z_i - z_{min} |$. Finally, please notice that if our landscape is completely flat, results expected by this extended model will be equal to those found by Chialvo and Millonas in [12], since $\Delta[i]/\Delta_{max}$ equals to zero. In this case, this is equivalent to say that only the swarm pheromonal field is affecting each ant choices, and not the *environment* - i.e. the expected network of trails depends largely on the initial random position of the colony, and in trail clusters formed in the

| | | |
|---|---|---|
| $t = 0$ | $t = 0$ | **Algorithm.** High-level description of the *SWARM SEARCH* algorithm proposed |
| $t = 500$ | $t = 500$ | |
| $t = 1000$ | $t = 1000$ | /* **Initialization** */ <br> **For** all agents **do** <br>   Place agent at randomly selected site <br> **End For** <br> /* **Main loop** */ <br> **For** $t = 1$ to $t_{max}$ **do** <br>   **For** all agents **do** <br>     /* **According to Eqs. 1 and 2 (section 3)** */ <br>     **Compute** $W(\sigma)$ and $P_{ik}$ <br>     **Move** to a selected neighboring site not occupied by other agent <br>     /* **According to Eq. 3 (section 3)** */ <br>     **Increase** pheromone at site $r$: <br>       $P_r = P_r + [\eta + p(\Delta[r]/\Delta max)]$ <br>   **End For** <br>   **Evaporate** pheromone by $K$, at all grid sites <br> **End For** <br> **Print** location of agents <br> **Print** pheromone distribution at all sites <br> /* **Values of parameters used in experiments** */ <br> $k = 0.015$, $\eta = 0.07$, $\beta = 3.5$, $\gamma = 0.2$, <br> $p = 1.9$, $t_{max} = 500, 600, 1000$ or $1150$ steps. <br> /* **Useful references** */ <br> Check [39], [41], [12], [33] and [32]. |
| $t = 1010$ | $t = 1010$ | |
| $t = 1050$ | $t = 1050$ | Fig.4 – **max$F0a$** => **max$F0b$**. Social evolution from maximizing function *F0a* to maximizing function *F0b*. In the first 1000 time steps the ant colony explores function *F0a*, while suddenly at $t=1001$, function *F0b* is used as the new *habitat*. Pheromone distribution (Social Cognitive Maps) for $t = 0$, 500, 1000, 1010, 1050, 1080, 1100 and 1150 time steps, of 3000 ants exploring function *F0a* and *F0b* on a 100 x 100 toroidal grid are shown. Already at $t=1010$, the old highest peak on the right suffers a radical erosion, on the presence of ants (they start to explore new regions). As time passes the majority of the colony moves to the new peak, on the left. The strategy pseudo-code is given above. |
| $t = 1080$ | $t = 1080$ | |
| $t = 1100$ | $t = 1100$ | |

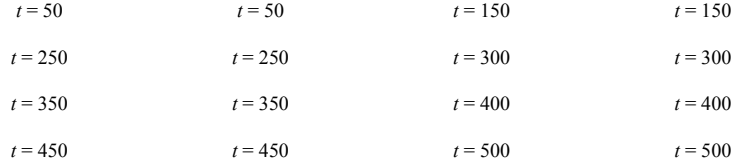

| | | | |
|---|---|---|---|
| *t* = 50 | *t* = 50 | *t* = 150 | *t* = 150 |
| *t* = 250 | *t* = 250 | *t* = 300 | *t* = 300 |
| *t* = 350 | *t* = 350 | *t* = 400 | *t* = 400 |
| *t* = 450 | *t* = 450 | *t* = 500 | *t* = 500 |

Fig.5 – **maxF0a => minF0a**. Maximizing function *F0a* during 250 time steps and then minimizing it for $t \geq 251$. Pheromone distribution (Social Cognitive Maps) for *t* = 50, 150, 250, 300, 350, 400, 450 and 500 time steps, of 2000 ants exploring function *F0a* on a 100 x 100 toroidal grid are shown. Already at *t*=300, the highest peak on the right suffers a radical erosion, on the presence of ants (they start to explore new regions). As time passes the majority of the colony moves to the deep valley, on the left. Parameters are different from those used in Figs. 3-4 (check table 2).

$$T = \eta + p \frac{\Delta[i]}{\Delta_{max}} \qquad (3)$$

initial configurations of pheromone. On the other hand, if this environmental term is added a stable and emergent configuration will appear which is largely independent on the initial conditions of the colony and becomes more and more dependent on the nature of the current studied *landscape* itself. As specified earlier in section 2.1, the environment plays an active role, in conjunction with continuous positive and negative feedbacks (section 2.2) provided by the colony and their pheromone, in order to achieve a stable emergent pattern, memory and distributed learning by the community.

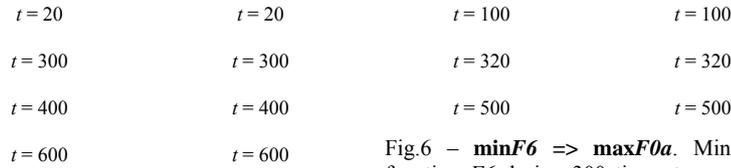

| | | | |
|---|---|---|---|
| *t* = 20 | *t* = 20 | *t* = 100 | *t* = 100 |
| *t* = 300 | *t* = 300 | *t* = 320 | *t* = 320 |
| *t* = 400 | *t* = 400 | *t* = 500 | *t* = 500 |
| *t* = 600 | *t* = 600 | | |

Fig.6 – **minF6 => maxF0a**. Minimizing function *F6* during 300 time steps and then maximizing function *F0a* for $t \geq 301$. Pheromone distribution (Social Cognitive Maps) for *t* = 20, 100, 300, 320, 400, 500, and 600 time steps, of 3000 ants exploring function *F6* and *F0a* on a 100 x 100 toroidal grid are shown. Parameters are different from those used in Figs. 3-4 (check table 2).

# 4 Experimental Setup and Results

In order to test the dynamical behaviour of this new *Swarm Search* algorithm presented earlier in section 3 (pseudo-code in fig. 4), we have used classical test functions (table 1) drawn from the literature in Genetic Algorithms, Evolutionary strategies and global optimization [38], several of them graphically accessible in fig. 2. Function *F0a* represents one deep valley and one peak, while *F0b* his opposite. Function *F1* represents *De Jong*'s function 1 and his one of the simplest. It is continuous, convex and unimodal; $x_i$ is in the interval [-5.12; 5.12] and the global minimum is at $x_i$=0. Function *F2* represents an axis parallel hyper-ellipsoid similar to *De Jong*'s function 1. It is also know as the weighted sphere model. Again it is continuous, convex and unimodal in the interval $x_i \rightarrow$ [-5.12; 5.12], with global minimum at $x_i$=0. Function *F3* represents an extension of the axis parallel hyper-ellipsoid (*F2*), also know as *Schwefel*'s function 1.2. With respect to the coordinate axes this function produces rotated hyper-ellipsoids; $x_i$ is in the interval [-65.536; 65.536] and the global minimum is at $x_i$=0. Likewise *F2*, it is continuous, convex and unimodal. Function *F4* represents the well-know *Rosenbrock*'s valley or *De Jong*'s function 2. *Rosenbrock*'s valley is a classic optimization problem. The global optimum is inside a long, narrow, parabolic shaped flat valley. To find the valley is trivial, however convergence to the global optimum is difficult and hence this problem has been repeatedly used in assess the performance of optimization algorithms; $x_i$ is in the interval [-2.048; 2.048] and the global minimum is at $x_i$=0. Function *F5* represents the *Rastrigin*'s function 6. This function is based on *De Jong*'s function 1 with the addition of cosine modulation to produce many local minima. Thus, the test function is highly multimodal. However, the location of the minima are regularly distributed. As in *F1*, $x_i$ is in the interval [-5.12; 5.12] and the global minimum is at $x_i$=0. Finally, *F6* represents *Schwefel*'s function 7, being deceptive in that the global minimum is geometrically distant, over the parameter space, from the next best local minima. Therefore, the search algorithms are potentially prone to convergence in the wrong direction; $x_i$ is in the interval [-500; 500] and the global minimum is at $x_i$=420,9687 while $f(x)=n.418,9829$. In our tests, $n$=2. Within this specific framework we have produced several run tests using different test functions, some of which are presented here trough figures 3 to 6. The parameters used are shown on table 2. The simplest test was the first one (fig.3) where we forced the colony to search for the maximal peak in function *F0a*, during 1000 time steps. The other tests were harder, since they include not only different purposes simultaneously (maximizing and minimizing) as well as different landscapes that changed dynamically on intermediate swarm search stages (e.g., fig. 6).

**Table 1.** Classical test functions used in our analysis from *MATLAB* reference manual [38]

| Function ID | Equation |
|---|---|
| F0a | $f_{0a}(x) = x_1 . e^{-0.2 \sum_{i=1}^{n} x_i^2}$ |
| F0b | $f_{0b}(x) = -x_1 . e^{-0.2 \sum_{i=1}^{n} x_i^2}$ |
| F1 | $f_1(x) = \sum_{i=1}^{n} x_i^2$ |
| F2 | $f_{1a}(x) = \sum_{i=1}^{n} i . x_i^2$ |
| F3 | $f_{1b}(x) = \sum_{i=1}^{n} \left( \sum_{j=1}^{i} x_j \right)^2$ |
| F4 | $f_2(x) = \sum_{i=1}^{n-1} 100 . (x_{i+1} - x_i^2)^2 + (1 - x_i)^2$ |
| F5 | $f_6(x) = 10.n + \sum_{i=1}^{n} (x_i^2 - 10 . \cos(2 . \pi . x_i))$ |
| F6 | $f_7(x) = \sum_{i=1}^{n} - x_i . \sin(\sqrt{|x_i|})$ |

**Table 2.** Parameters used for different test sets (check section 3 and figures 3-6 for details)

| Test | Fig. | $N$ ants | $t_{max}$ | $k$ | $\eta$ | $\beta$ | $\gamma$ | $p$ |
|---|---|---|---|---|---|---|---|---|
| **max*F0a*** | 3 | 3000 | 1000 | 0.015 | 0.07 | 3.5 | 0.2 | 1.93 |
| **max*F0a* => max*F0b*** | 4 | 3000 | 1150 | 0.015 | 0.07 | 3.5 | 0.2 | 1.93 |
| **max*F0a* => min*F0a*** | 5 | 2000 | 500 | 1.000 | 0.10 | 3.5 | 0.2 | 1.90 |
| **min*F6* => max*F0a*** | 6 | 3000 | 600 | 1.000 | 0.01 | 3.5 | 0.2 | 1.90 |

### 4.1 SWARM SEARCH *versus* BACTERIAL FORAGING algorithms

In order to further analyze the collective behavior of the present proposal, we make a comparison between the ant-like *Swarm Search Algorithm* (SSA) and the *Bacterial Foraging Optimization Algorithm* (BFOA), on the dominion of function optimization. BFOA was selected since it represents an earlier proposal for function optimization as well based on natural foraging capacities. Presented by Passino at *IEEE Control Systems Magazine* in 2002 [36] and later that year in the *Journal of Optimization Theory and Applications* [28], the author for the purpose of a simple but powerful illustrative example, used his algorithm to find the minimum of two complex functions $J_{cc}$, described in [36], page 60. Further material, as the MATLAB code of his algorithm and the tri-dimensional functions experimented, can also be found on the web address of a recent book from the same author (*Biomimicry for Optimization, Control and Auto-*

*mation*, Springer-Verlag, London, UK, 2005), at *http://www.ece.osu.edu/~passino/ICbook/ ic_index.html*. Passino uses $S$=50 bacteria-based agents, during four generations. In each generation, and has a requirement of his algorithm, each agent enters a chemotaxis loop (see page 61 [36]), performing $N_c$=100 chemotactic (foraging) steps. Thus the algorithm – for the precise application – runs for $t$=400 time steps, which make us believe that a fair comparison can be make in regard of the parameter values we use. The two functions represent what Passino designates by *nutrient concentration landscapes* (see fig. 7, first row – the web address also contains his MATLAB code used in the two functions, where *Nutrientsfunc.m* and *Nutrientsfunc1.m* are represented by different weights). His function *F2* (*Nutrientsfunc1.m*) has a zero value at [15,15] and decreases to successively more negative values as you move away from that point, reaching a *plateau* with the same value. Moreover, and for the purpose of discrete function optimization, Passino [36] represented both functions by a discrete lattice (as well as us in our past tests) with a size of 30 x 30 cells over the optimization domain (each cell has a correspondent $z$ or $J_{cc}$ value). For these reasons and in order to keep a coherent comparison, we shall use 50 ant-like agents in our SSA, on a 30 x 30 tri-dimensional *habitat*, for $t$=400 time steps, on both functions. We then run 3 tests. The first is requested to minimize Passino's function *F1*. The second test is requested to minimize Passino's function 2. Finally, and in order to prove the adaptive features of our model, we requested SSA to deal with two contradictory goals, i.e. to minimize *F1* and then to maximize it, over the same period of 400 time steps. As visible, SSA quickly adapts to the different purposes. Over function *F1*, the pheromone concentration is already intensely allocated at the right point at $t$=100 (and not in other areas), while BFOA still explores different regions on the optimization domain. Over function *F2*, the swarm quickly separates in different foraging groups, since there are a large number of points with the minimal value. Finally over function *F1* again, in the final test (last row – fig. 7), SSA is able to process two different demands (maximization followed by minimization) over the same foraging time period that BFOA uses for *F1* minimization. The parameters used in our experiments follows: $N_{ants}$=50, $t_{max}$=400, $k$ =1 (pheromone evaporation rate), $\eta$=0.1 (pheromone deposition rate), $\beta$=7 (this parameter controls how ants follow the pheromone gradient), $\gamma$=0.2, and $p$=1.9. In test 1, $\beta$=6.

| *Passino F1 – 3D* | *Passino F1 – 2D* | *Passino F2 – 3D* | *Passino F2 – 2D* |
|---|---|---|---|
| $t = 100$ | $t = 200$ | $t = 300$ | $t = 400$ |
| $t = 100$ | $t = 200$ | $t = 300$ | $t = 400$ |
| $t = 100$ | $t = 200$ | $t = 300$ | $t = 400$ |
| $t = 100$ | $t = 200$ | $t = 300$ | $t = 400$ |
| $t = 100$ | $t = 200$ | $t = 300$ | $t = 400$ |

Fig.7 – In the first row the test functions used by Passino [36,28]. In the second and third rows, BFOA minimizing results respectively for *F1* and *F2*. The graphics show the bacterial motion trajectories (using 50 bacteria-like agents). In the fourth and fifth rows, SWARM-SEARCH algorithm (SSA) minimizing results respectively for *F1* and *F2*, and for the same foraging time period. The graphics shows the pheromone distribution. In the last row, SSA is requested to deal with two contradictory goals, i.e. to minimize *F1* and then to maximize it. In all these tests, SSA has used 50 ant-like agents. Check main text for the parameters used. Habitat size equals 2 x [0,30].

## 5 Conclusions

Evolution of mass behaviours on time are difficult to predict, since the global behaviour is the result of many part relations operating in their own local neighbourhood. The emergence of network trails in ant colonies, for instance, are the *product* of several simple and local interactions that can evolve to complex patterns, which in some sense translate a meta-behaviour of that swarm. Moreover, the translation of one kind of low-level (present in a large number) to one meta-level is minimal. Although that behaviour is specified (and somehow constrained), there is minimal specification of the mechanism required to generate that behaviour; global behaviour evolves from the many relations of multiple simple behaviours, without global coordination (i.e. from local interactions to global complexity). One paradigmatic and abstract example is the notion, within a specified population, of *common-sense*, being the meta-result a type of *collective-conscience*. There is some evidence that our brain as well as many other complex systems, operates in the same way, and as a consequence collective perception capabilities could be derived from emergent properties, which cannot be neglected in any pattern search algorithm. These systems show in general, interesting and desirable features as *flexibility* (e.g. the brain is able to cope with incorrect, ambiguous or distorted information, or even to deal with unforeseen or new situations without showing abrupt performance breakdown) or *versability*, *robustness* (keep functioning even when some parts are locally damaged), and they operate in a massively parallel fashion. Present results point to that type of interesting features. Although the current model is far from being consistent with real ones, since only some type of mechanisms are considered, swarm pheromonal fields reflect some convergence towards *the identification* of a common goal in a purely decentralized from. Moreover, the present model shows important adaptive capabilities, as in the presence of sudden changes in the *habitat* - our test landscapes (fig. 3). Even if the model is able to quickly adapt to one specific environment, evolving from one empty pheromonal field, *habitat* transitions point that, the whole system is able to have some memory from past environments (i.e. convergence is more difficult after *learning* and *perceiving* one past *habitat*). This emerged feature of *résistance*, is somewhat present in many of the natural phenomena that we find today in our society. In a certain sense, the distribution of pheromone represents the collective solutions found so far (memory, risk avoidance, exploitation behavior), while evaporation enables the system to adapt (tricks a decision, explorative behavior), not only as in normal situations (a complex search landscape), as well as when the landscape suddenly changes, moving the colony's new target to a new unexplored region. And again as noted by Langton [27,26], as in many complex systems, only at the right intermediary regime, in here between contradictory behaviors of exploration and exploitation, the swarm is able to quickly converge.

The recognizable results indicate that the collective intelligence is able to cope and quickly adapt to unforeseen situations even when over the same cooperative foraging period, the community is requested to deal with two different and contradictory purposes. All these above mentioned aspects show how vital can be the study of social foraging for the development of new distributed search algorithms, and the construction of social cognitive maps, with interesting properties in collective memory, collective decision-making and swarm pattern detection.

But the work could have important consequences in other areas. Perhaps, one of the most valuable relations to explore is that of social foraging and evolution. For two reasons; First, as described by Passino [36], natural selection tends to eliminate animals with poor "foraging strategies" (methods for locating, handling, and ingesting food) and favor the propagation of genes of those animals that have successful foraging strategies since they are more likely to enjoy reproductive success (they obtain enough food to enable them to reproduce). Logically, such evolutionary principles have led scientists in the field of foraging theory to hypothesize that it is appropriate to model the activity of foraging as an optimization process: A foraging animal takes actions to maximize the energy obtained per unit time spent foraging, in the face of constraints presented by its own physiology and by the environment.

Second, because there is an increasing recognition that natural selection and self-organization work hand in hand to form evolution, as defended by Kauffmann [22,23]. For example, anthropologist Jeffrey McKee [31,24] has described the evolution of human brain as a self-organizing process. He uses the term autocatalysis to describe how the design of an organism's features at one point in time affects or even determines the kinds of designs it can change into later. For example the angle of the skull on the top of the spine left some extra space for the brain to expand. Thus the evolution of the organism is determined not only by selection pressures but by constraints and opportunities offered by the structures that have evolved so far. Also, and back again in what regards the evolution of collectives, it is known that during the evolution of life, there have been several transitions in which individuals began to cooperate, forming higher levels of organization and sometimes losing their independent reproductive identity (insect societies are one example). Several factors that confer evolutionary advantages on higher levels of organization have been proposed, such as *Division of Labor* and *Increased Size*. But recently, a new third factor was added: *Information Sharing* [25]. Lachmann et al, illustrate with a simple model how information sharing can result in individuals that both receive more information about their environment and pay less for it.

Being social foraging essentially a self-organized phenomenon, the study of computational foraging embedded with GA (Genetic Algorithm) like natural selection can much probably enhance our understanding on the detailed forms of the hypothetical equation: *Evolution = Natural Selection + Self-Organization*, and in the precise role of each "variable". As an example, current work by present authors, include the research of variable population size swarms, as used similarly in Evolutionary Computation (Fernandes and Rosa [17]), where each individual can have a probability of making a child, as well to die, depending on his *accumulated* versus *spent* energetic resources. The system as a whole, then proceeds on the search space as a kind of distributed evolutionary swarm.

**References**

1. Bak, P., *How Nature Works – The Science of Self-Organized Criticality*, Springer-Verlag, 1996.
2. Ball, P., "Do Plants act like Computers?", Nature News Service, *MacMillan* Magazines Ltd, 21 Jan. 2004.
3. Bassler, B.L, "Small Talk: Cell-to-Cell Communication in Bacteria", *Cell*, Vol. 109, pp. 421-424, May 2002.
4. Ben-Jacob, E., Shochet, O., Tenenbaum, A., Cohen, I., Czirók, A., Vicsek, T., "Generic Modelling of Cooperative Growth in Bacterial Colonies", *Nature*, 368, pp. 46-49, 1994.
5. Ben-Jacob, E., "Bacterial Wisdom, Gödel's Theorem and Creative Genomic Webs", *Physica A*, 248, pp. 57-76, 1998.
6. Ben-Jacob, E., Becker, I., Shapira, Y., Levine, H., "Bacterial Linguistic Communication and Social Intelligence", *Trends in Microbiology*, Vol. 12/8, pp. 366-372, 2004.
7. Bonabeau, E., Theralauz, G., Deneubourg, J.-L., Camazine S., "Self-Organization in Social Insects", *Trends in Ecology and Evolution*, 12(5), 188-193, 1997.
8. Bonabeau, E., Dorigo, M., Theraulaz, G., *Swarm Intelligence: From Natural to Artificial Systems*, Santa Fe Institute in the Sciences of Complexity, Oxford Univ. Press, New York, Oxford, 1999.
9. Camazine, S., Deneubourg, J.-L., Franks, N.R., Sneyd, J., Theraulaz, G., Bonabeau, E., *Self-Organization in Biological Systems*, Princeton Studies in Complexity, Princeton University Press, Princeton and Oxford, 2001.
10. Castro, L.N., Timmis, J.I., *Artificial Immune Systems: A New Computational Intelligence Approach*, Springer-Verlag, 2002.
11. Castro, L.N., Von Zuben, F.J., *Recent Developments in Biologically Inspired Computing*, Eds., Idea Group Publishing Inc., 2005.
12. Chialvo, D.R., Millonas, M.M., "How Swarms build Cognitive Maps", In Steels, L. (Ed.): *The Biology and Technology of Intelligent Autonomous Agents*, 144, NATO ASI Series, 439-450, 1995.
13. Chialvo, D.R., "Critical Brain Networks", *Physica A*, 340, 4, 756, 2004.
14. Chowdhury, D., Nishinari, K., Schadschneider, A., "Self-Organized Patterns and Traffic Flow in Colonies of Organisms: from Bacteria and Social Insects to Vertebrates", special issue on Pattern Formation, in *Phase Transitions*, Taylor and Francis, vol. 77, 601, 2004.
15. Dasgupta, D., *Artificial Immune Systems and Their Applications*, Ed., Springer-Verlag, 1999.
16. Farmer, J.D., "A Rosetta Stone for Connectionism", in S. Forrest (Ed.), *Emergent Computation*, Cambridge, MA: The MIT Press, 1991.
17. Fernandes, C., Rosa, A.C., "Study on Non-random Mating and Varying Population Size in Genetic Algorithms using a Royal Road Function", IEEE CEC´01, Proc. of the 2001 IEEE Congress on Evolutionary Computation, pp. 60-66, 2001.
18. Grilo, A., Caetano, A., Rosa, A.C., "Agent based Artificial Immune System", Proc. GECCO-01, Vol. LBP, pp. 145-151, 2001.
19. Hofstadter, D.R., *Gödel, Escher, Bach: An Eternal Golden Braid*, New York: Basic Books, 1979.

20. Holland, O., Melhuish, C.: Stigmergy, "Self-Organization and Sorting in Collective Robotics", *Artificial Life*, Vol. 5, n. 2, MIT Press, 173-202, 1999.
21. Jeanne, R.L., "The Evolution of the Organization of Work in Social Insects", *Monit. Zool. Ital.*, 20, 119.133, 1986.
22. Kauffmann, S.A., *The Origins of Order: Self-Organization and Selection in Evolution*, New York: Oxford University Press, 1993.
23. Kauffmann, S.A., At *Home in the Universe: The Search for the Laws of Self-Organization and Complexity*, New York: Oxford University Press, 1995.
24. Kennedy, J. Eberhart, Russel C. and Shi, Y., *Swarm Intelligence*, Academic Press, Morgan Kaufmann Publ., San Diego, London, 2001.
25. Lachmann, M., Sella, G., Jablonka, E., "On Information Sharing and the Evolution of Collectives", Proc. of the Royal Society: *Biological Sciences*, 267, pp. 1265-1374, 2000.
26. Langton, C.G., "Computation at the Edge of Chaos", *Physica D*, 42, pp. 12-37, 1990.
27. Langton, C.G., "Life at the Edge of Chaos", in C.G. Langton (Ed.), *Artificial Life* II, Santa Fe Institute Studies in the Sciences of Complexity, Vol. 10, Reading, MA: Addison-Wesely, pp. 41-91, 1991.
28. Liu, Y., Passino, K.M., "Biomimicry of Social Foraging Bacteria for Distributed Optimization: Models, Principles, and Emergent Behaviors", Journal of Optimization Theory and Applications, Vol. 115, n°3, pp. 603-628, Dec. 2002.
29. Liu, Y., Passino, K.M., "Stable Social Foraging Swarms in a Noisy Environment," *IEEE Trans. on Automatic Control*, Vol. 49, No. 1, pp. 30-44, Jan. 2004.
30. Maree, A.F.M., Hogeweg, P., "How Amoeboids Self-Organize into a Fruiting Body: Multicellullar Coordination in *Dictyostelium discoideum*", PNAS, vol. 98, n° 7, pp. 3879-3883, 2001.
31. McKee, J.K., *The Riddled Chain: Change, Coincidence, and Chaos in Human Evolution*, Piscataway, NJ: Rutjers University Press, 2000.
32. Millonas, M.M., "A Connectionist-type model of Self-Organized Foraging and Emergent Behavior in Ant Swarms", *J. Theor. Biol.*, n° 159, 529, 1992.
33. Millonas, M.M., "Swarms, Phase Transitions and Collective Intelligence", In Langton, C.G. (Ed.): *Artificial Life III*, Santa Fe Institute, Studies in the Sciences of Complexity, Vol. XVII, Addison-Wesley, Reading, Massachusetts, 417-445, 1994.
34. Panait, L., Luke, S., "Evolving Foraging Behaviors", Proc. of the Ninth Int. Conf. on the Simulation and Synthesis of Living Systems, ALIFE-9, 2003.
35. Panait, L., Luke, S., "A Pheromone-based Utility Model for Collaborative Foraging", Proc. of the 2004 Conference on Autonomous Agents and Multiagent Systems, 2004.
36. Passino, K.M., "Biomimicry of Bacterial Foraging for Distributed Optimization and Control", *IEEE Control Systems Magazine*, pp. 52-67, June 2002.
37. Peak, D.A., West, J.D., Messinger, S.M., Mott, K.A., "Evidence for Complex, Collective Dynamics and Emergent, Distributed Computation in Plants", *PNAS*, Proc. of the National Academy of Sciences, USA, 101, pp. 918-922, 2004.
38. Pohlheim, H, "Genetic Algorithm MATLAB Toolbox Test Functions", MATLAB reference manual, version 1.2, *Mathworks*, 1997.
39. Ramos, V., Almeida, F., "Artificial Ant Colonies in Digital Image Habitats: A Mass Behavior Effect Study on Pattern Recognition", In Dorigo, M., Middendorf, M., Stuzle, T. (Eds.): *From Ant Colonies to Artificial Ants* - 2$^{nd}$ Int. Wkshp on Ant Algorithms, 113-116, 2000.
40. Ramos, V., "On the Implicit and on the Artificial - Morphogenesis and Emergent Aesthetics in Autonomous Collective Systems", in ARCHITOPIA Book, *Art, Architecture and Science*, Institut D'Art Contemporain, J.L. Maubant et al. (Eds.), pp. 25-57, Chapter 2, ISBN 2905985631 – EAN 9782905985637, France, Feb. 2002.
41. Ramos,V., Merelo, Juan J., "Self-Organized Stigmergic Document Maps: Environment as a Mechanism for Context Learning", in AEB'2002 – *1$^{st}$ Spanish Conf. on Evolutionary and*

*Bio-Inspired Algorithms*, E. Alba, F. Herrera, J.J. Merelo et al. (Eds.), pp. 284-293, Centro Univ. de Mérida, Mérida, Spain, 6-8 Feb. 2002.
42. Ramos, V., Abraham, A., **"**Evolving a Stigmergic Self-Organized Data-Mining", in ISDA-04, *4$^{th}$ Int. Conf. on Intelligent Systems, Design and Applications*, Budapest, Hungary, ISBN 963-7154-30-2, pp. 725-730, August 26-28, 2004.
43. Robinson, G.E., "Regulation of Vision of Labor in Insect Societies", *Annu. Rev. Entomol.*, 37, 637-665, 1992.
44. Savill, N.J., Hogeweg, P., "Modelling Morphogenesis: From Single Cells to Crawling Slugs", J. Theor. Biology, 184, pp. 229-235, 1997.
45. Seeley, T.D., "The Honey Bee Colony as a Super-Organism", *American Scientist*, 77, pp. 546-553, 1989.
46. Sierakowski, C.A., Coelho, L.S., "Path Planning Optimization for Mobile Robots based on Bacteria Colony Approach" (in Portuguese), VIII Simpósio Brasileiro de Redes Neurais, 2004.
47. Silberman, S., "The Bacteria Whisperer", *Wired*, 11-04, pp. 104-108, April 2003.
48. Sporns, O., Chialvo, D.R., Kaiser, M., Hilgetag, C.C., "Organization, Development and Function of Complex Brain Networks", *Trends in Cognitive Sciences*, Vol. 8, nº9, Sept. 2004.
49. Strogatz, S., *Sync – The Emerging Science of Spontaneous Order*, Theia, New York, 2003.
50. Theraulaz, G., Bonabeau, E., "Coordination in Distributed Building", *Science*, 269, 686-688, 1995.
51. Theraulaz, G., Bonabeau, E., "Modelling the Collective Building of Complex Architectures in Social Sciences wit Lattice Swarms", *J. Theor. Biol.*, 1777, 381-400, 1995.
52. Theraulaz, G., Bonabeau, E., "A Brief History of Stigmergy", *Artificial Life*, Vol. 5, n. 2, MIT Press, 97-116, 1999.
53. Williams, H., *Spatial Organisation of a Homogeneous Agent Population using Diffusive Signalling and Role Differentiation*, MSc Evolutionary & Adaptive Systems Thesis, COGS, Univ. of Sussex, UK, Sept. 2002.
54. Wilson, E.O., *The Insect Societies*, Cambridge, MA., Belknap Press, 1971.
55. Wolfram, S., "Universality and Complexity in Cellular Automata", *Physica D*, 10, pp. 1-35, 1984.